\documentstyle[twoside,fleqn,photon99,amssymb]{article}

\def\gp{\gamma p}
\def\ga{\gamma\gamma}
\def\ee{e^+e^-}
\def\etjet{E_T^{jet}}
\def\etajet{\eta^{jet}}
\def\phijet{\varphi^{jet}}
\def\rs{R_{SEP}}
\def\kt{k_T}
\def\rr1{R=1}
\def\r7{R=0.7}
\def\etaphi{\eta-\varphi}
\def\q2{Q^2}
\def\g2{GeV$^2$}
\def\xo{x_{\gamma}^{OBS}}
\def\f2{F_2^{\gamma}}
\def\mj{M^{JJ}}
\def\cost{\vert\cos\theta^*\vert}

\def\scost{d\sigma/d\cost}
\def\seta{d\sigma/d\etajet}
\def\spt{d\sigma/dp_T}

\def\Journal#1#2#3#4{{#1} {\bf #2} (#4) #3}

\def\NPB{{\em Nucl. Phys.} B}
\def\PLB{{\em Phys. Lett.}  B}
\def\PRL{{\em Phys. Rev. Lett.}}
\def\PRD{{\em Phys. Rev.} D}
\def\PRX{{\em Phys. Rev.}}
\def\PRP{{\em Phys. Rep.}}
\def\ZPC{{\em Z. Phys.} C}
\def\ZPX{{\em Z. Phys.}}
\def\EPC{{\em Eur. Phys. Jour.} C}
\def\CPC{{\em Comp. Phys. Comm.}}

\title{Inclusive hard processes in $\ga$ and $\gp$ interactions$^\dag$}

\author{C. Glasman\address{Departamento de F\'\i sica Te\'orica C-XI,
Facultad de Ciencias, Universidad Aut\'onoma de Madrid,\\
Cantoblanco, 28049 Madrid, Spain}
\thanks{Supported by an EC fellowship number ERBFMBICT 972523.
\hspace{6cm}
$^\dag$Invited talk at the {\it International Conference on the Structure
and Interactions of the Photon} (PHOTON 99), Freiburg im Breisgau,
Germany, May $23^{rd}-27^{th}$, 1999.
}}
       
\begin{document}

\begin{abstract}
Measurements of jet, prompt photon, high$-p_T$ hadron and heavy quark
production in photon-induced processes provide tests of QCD and are sensitive
to the photon parton densities. A review of the latest experimental results in
$\ga$ and $\gp$ interactions is presented. Next-to-leading-order QCD
calculations for these measurements are discussed.
\end{abstract}

\maketitle

\setcounter{footnote}{0}
\section{Introduction}
The structure of the photon can be studied via inclusive hard processes
in photon-photon interactions in $\ee$ collisions and in photon-proton
interactions in $ep$ collisions. At LEP, centre-of-mass energies ranging from
161 to 172 GeV allow to probe virtualities of the photon in the range
$0\lesssim\q2\lesssim 10^4$~\g2\ and $\ga$ centre-of-mass energies in the
range $10<W_{\ga}<110$~GeV. At HERA, electrons (positrons) of energy
$E_e=27.5$~GeV and protons of energy $E_p=820$~GeV collide at a centre-of-mass
energy of 300~GeV. The virtuality of the photon ranges within
$0\lesssim\q2\lesssim 10^5$~\g2\ and the $\gp$ centre-of-mass energy varies in
the range $130<W_{\gp}<280$~GeV.

Tests of QCD and information on the parton densities in the photon and
in the proton are obtained from measurements of jet, prompt photon, high$-p_T$
hadron and heavy quark production. The structure of the real and virtual
photon is measured directly in deep inelastic $e\gamma$ scattering
(DIS $e\gamma$). Measurements of jet substructure and hadron production allow
the study of parton radiation and fragmentation.

\section{Jet cross sections}
At leading order (LO) QCD, three processes contribute to jet production
in $\ga$ interactions: in the double-resolved (DR) processes a parton
from one of the photons interacts with a parton from the other photon to
give two jets in the final state; in the single-resolved (SR) processes,
one of the photons interacts as a point-like particle with a parton from
the other photon; and in the direct (DD) processes, both photons
interact as point-like particles.

In $\gp$ interactions, there are only two processes which contribute to
jet production at LO QCD: the resolved (R) processes, in which the photon
interacts through its partonic content with a parton from the proton,
and the direct (D) processes, in which the photon interacts as a
point-like particle.

The cross section for jet production at LO QCD in $\gp$ interactions is
given by

$$\sigma^{LO,ep}_{D} = \int d\Omega\ f_{\gamma /e}(y)\ 
f_{j/p}(x_p,\mu^2_F)\times$$
\rightline{$d\sigma(\gamma j\rightarrow {\rm jet}\ {\rm jet})$}
and
$$\sigma^{LO,ep}_{R} = \int d\Omega\ f_{\gamma /e}(y)\ 
f_{i/\gamma}(x_{\gamma},\mu^2_F)\times$$
\rightline{$f_{j/p}(x_p,\mu^2_F)\
d\sigma(ij\rightarrow {\rm jet}\ {\rm jet}),$}

\vspace{0.5cm}
\parindent 0em
where $f_{\gamma /e}(y)$ is the flux of photons in the
electron~\footnote{The variable $y$ is the fraction of the electron
energy taken by the photon.} which is usually estimated by the
Weizs\"acker-Williams approximation~\cite{wwa}; $f_{j/p}(x_p,\mu^2_F)$
are the parton densities in the proton (determined from e.g. global
fits~\cite{mrs}), $x_p$ is the fraction of the proton momentum taken by
parton $j$ and $\mu_F$ is the factorisation scale; and
$d\sigma(\gamma(i)j\rightarrow {\rm jet}\ {\rm jet})$ is the subprocess
cross section, calculable in perturbative QCD (pQCD). In the case of resolved
processes, there is an additional ingredient:
$f_{i/\gamma}(x_{\gamma},\mu^2_F)$ are the parton densities in the photon, for
which there is only partial information, and $x_{\gamma}$ is the fraction of
the photon momentum taken by parton $i$. The integrals are performed over the
phase space represented by ``$d\Omega$''.
\parindent 1em

For $\ga$ interactions, the contributions to the jet production cross
section at LO QCD may be written as

$$\sigma^{LO,\ee}_{DD} = \int d\Omega\ f_{\gamma /e}(y^{(1)})\ 
f_{\gamma /e}(y^{(2)})\times$$
\rightline{$d\sigma(\gamma\gamma\rightarrow {\rm jet}\ {\rm jet}),$}

$$\sigma^{LO,\ee}_{SR} = \int d\Omega\ f_{\gamma /e}(y^{(1)})\ 
f_{\gamma /e}(y^{(2)})\times$$
\rightline{$f_{i/\gamma}(x_{\gamma}^{(1)},\mu^2_F)\
d\sigma(i\gamma\rightarrow {\rm jet}\ {\rm jet}),$}
and
$$\sigma^{LO,\ee}_{DR} = \int d\Omega\ f_{\gamma /e}(y^{(1)})\ 
f_{\gamma /e}(y^{(2)})\times$$
\rightline{$f_{i/\gamma}(x_{\gamma}^{(1)},\mu^2_F)\
f_{j/\gamma}(x_{\gamma}^{(2)},\mu^2_F)\
d\sigma(ij\rightarrow {\rm jet}\ {\rm jet}).$}

\vspace{0.5cm}
The separation of resolved and direct processes used above has a meaning only
at LO. For $\gp$ interactions, the variable $\xo$,

$$
\xo={1\over 2yE_e}({E_T^{jet1}e^{-\eta^{jet1}}+E_T^{jet2}e^{-\eta^{jet2}}}),
$$
gives the fraction of the photon energy invested in the production of the
dijet system. $E_T^{jet1(2)}$ and $\eta^{jet1(2)}$ are the transverse energy
and pseudorapidity\footnote{The pseudorapidity is defined as
$\eta=-\ln(\tan\frac{\theta}{2})$, where $\theta$ is the polar angle.} of the
two highest transverse energy jets in the event, respectively. The $\xo$
variable can be used to define resolved and direct processes in a meaningful
way to all orders. At LO QCD, $\xo=x_{\gamma}=1$ for direct processes.

At LO QCD, resolved processes dominate the jet production cross section for
low $\etjet$ and high $\etajet$ and direct processes dominate for high
$\etjet$ and low $\etajet$ (see e.g. \cite{klasen1}).

The calculations of jet cross sections in pQCD depend on the factorisation
($\mu_F$) and the renormalisation ($\mu_R$) scales. At LO, the dependence on
the factorisation scale comes from the resolved part and is very large. It
is largely canceled by the next-to-leading order (NLO) QCD direct
contribution and the factorisation scale dependence of the full NLO
calculation is small. This is the reason why the separation of resolved and
direct processes in terms of the Feynman diagrams is unphysical beyond LO.
Also, LO QCD calculations present a large dependence on the renormalisation
scale, which is reduced at NLO.

\subsection{NLO QCD corrections}
Jet cross sections are calculated by applying a jet algorithm to the
partons in the final state. The result of this procedure is trivial at LO, and
only at NLO the cross section depends on the jet algorithm, e.g. a cone jet
algorithm (see section 2.2). The NLO corrections to the jet cross section,
$$\sigma^{NLO}=\int d\Omega\ \{d\sigma^{LO}\ +\ 
{\alpha_s(\mu_R)\over 2\pi} K_{ij}(R,\mu_F,\mu_R)\}$$
depend on the radius of the cone jet algorithm $R$, and the factorisation and
the renormalisation scales.

There are several complete calculations of jet cross sections at NLO for $\ga$
and $\gp$ interactions \cite{klasen1,harris,frixione,aurenche}. Two types of
corrections contribute at NLO: the virtual corrections which include internal
particle loops and the real corrections which include a third parton
in the final state. The existing calculations differ mainly in the
treatment of the real corrections. In ref. \cite{klasen1} the phase
space slicing method is used with an invariant mass cut to isolate the
singular regions of the phase space. In ref. \cite{harris}, the phase
space slicing method is also used but with two small cut-offs to
delineate the soft and collinear regions. On the other hand, ref.
\cite{frixione} uses the subtraction method. With these methods, no
unphysical cut-offs parameters are introduced and the dependence on the
renormalisation scheme and the renormalisation and factorisation scales
is reduced. These methods also allow the implementation of jet
definitions.

A detailed comparison of the calculations from \cite{klasen1},
\cite{harris} and \cite{frixione} has been carried out in \cite{khf}
for NLO calculations of dijet cross sections and LO calculations of
three-jet cross sections. A reasonable agreement between the different
calculations was found, and the differences are up to $\sim 5\%$.

\subsection{The iterative cone algorithm}
In hadronic type interactions, jets are usually reconstructed by a cone
algorithm \cite{cone}. In NLO calculations, jet algorithms are applied
on the partons and since there are only three in the final state, only
up to two partons may build up a jet. Experimentally, jets are found in
the pseudorapidity ($\eta$) $-$ azimuth ($\varphi$) plane using the
transverse energy flow of the event. The jet variables are defined
according to the Snowmass Convention \cite{snow},

$$\etjet = \sum_i E^i_T$$
$$\etajet = \frac{\sum_i E^i_T\cdot\eta_i}{\etjet};\
\phijet = \frac{\sum_i E^i_T \cdot\varphi_i}{\etjet}.$$

In the iterative cone algorithm, jets are searched by maximising the
summed transverse energy within a cone of radius $R$.
It may happen that jets overlap and an additional rule is needed to decide
whether the jets should be merged. Given the limitations in the NLO
calculations, the overlapping of jets in the data cannot be reproduced. This
effect leads to uncertainties in the comparison between data and theory.

\subsection{Comparing theory and experiment}
To perform tests of QCD and to extract information on the photon parton
densities, the experimental and theoretical uncertainties should be
reduced as much as possible.

Among the main experimental uncertainties we find the presence of a
possible underlying event, which is the result of soft interactions between the
partons in the photon and proton remnants (for resolved events), and is not
included in the calculations. The uncertainty in the absolute energy scale of
the jets is usually the major source of experimental uncertainty. Jet cross
sections are usually measured at the hadron level, whereas the predictions are
calculated at the parton level. Therefore, when comparing theory and
experiment, hadronisation effects become another source of uncertainty.

On the theoretical side, since calculations are made only at NLO, the
implementation of the iterative cone jet algorithm in the theory does not
match exactly the experimental procedure. Furthermore, the uncertainty due to
the missing higher orders QCD terms may be large.

\subsubsection{The underlying event}
The presence of a possible underlying event has been investigated by
the OPAL and H1 collaborations by studying the transverse energy flow outside
the cone of the jets as a function of $\xo$ \cite{opal1,h11}. For low values
of $\xo$, the leading-logarithm parton-shower Monte Carlo calculations predict
a smaller transverse energy flow than in the data when no multipartonic
interactions (MI) are added. The multipartonic interactions are used to
simulate the effect of the extra energy inside and outside of the jet cone
that the soft underlying event would produce.

The ZEUS collaboration has studied the transverse energy flow around the jet
axis as a function of $\etajet$ \cite{zeus1}. The jets in the data are broader
at large $\etajet$ than in the Monte Carlo without MI. In particular, an
excess of transverse energy towards the proton direction is observed in the 
data with respect to the Monte Carlo calculations.

All these measurements give an indication of the presence of a possible
underlying event for low values of $\xo$, high values of $\etajet$ and
low $\etjet$. It is not possible to subtract unambiguously the contribution
from the underlying event since a fraction of the transverse energy flow
outside the jets comes from QCD radiation. A better solution is to find a
region of phase space where the contribution of the underlying event is
expected to be smaller.

To investigate the experimental uncertainties due to the underlying event, the
ZEUS collaboration has made measurements of $\seta$ using the iterative cone
algorithm with different cone radii and different $\etjet$ thresholds
\cite{zeus2}. NLO calculations \cite{klasen1,harris} have been compared to the
measurements: for $\rr1$, the data show an excess with respect to the
calculations at high $\etajet$ for low $\etjet$ thresholds. This excess
disappears if the cone radius is reduced to $\r7$ or the $\etjet$ threshold is
increased to 21~GeV; in both cases the NLO calculations provide a good
description of the data in the whole $\etajet$ range studied.

\subsubsection{The energy scale and hadronisation uncertainties}
The uncertainty on the absolute energy scale of the jets is usually
the main source of experimental uncertainty. For the ZEUS experiment,
this uncertainty is estimated to be $\pm 3\%$ \cite{zeus2}. The effect
on the jet cross sections is $\approx 12\%$ and is highly correlated
between measurements at different points. The H1 collaboration estimates
the uncertainty on the energy scale to be $\pm 4\%$ \cite{h12} which
amounts to $\approx 20\%$ in the jet cross sections. For the OPAL
experiment, the uncertainty on the energy scale is estimated to be $\pm 5\%$
\cite{opal1}.

NLO QCD calculations refer to jets built out of at most two partons whereas
the measurements refer to jets at the hadron level. An estimate of the
effects of hadronisation has been obtained by ZEUS \cite{zeus2} by
comparing the cross sections for jets of hadrons and jets of partons
calculated using the leading-logarithm parton-shower Monte Carlo PYTHIA
\cite{pythia} and it amounts to $10\ (20)\%$ for $\rr1\ (\r7)$. In
PYTHIA, fragmentation into hadrons is performed using the LUND
\cite{lund} string model as implemented in JETSET \cite{jetset}.
Using similar methods, H1 and OPAL estimated this uncertainty to be
$20\%$ \cite{h12} and $20-30\%$ \cite{opal1}, respectively, for their
measurements.

\subsubsection{Theoretical uncertainties in NLO QCD calculations}
The uncertainty coming from the missing higher-order QCD terms in the
NLO calculations is usually estimated by varying the renormalisation
and factorisation scales from $\etjet/4$ to $\etjet$. The effect depends
on $\etjet$ and $R$. For the phase space region in \cite{zeus2}, it
amounts to $\sim 5\ (20)\%$ for $\r7\ (\rr1)$ and $\etjet>21$~GeV.

The uncertainty in matching the experimental and theoretical jet
algorithms is particularly important for the iterative cone algorithm
due to the effects of overlapping and merging of jets. A partial
solution is the introduction of an ad-hoc parameter called $\rs$
\cite{ellis} to control parton recombination in the theoretical jet
algorithm according to the recipe that two partons are not combined if
their distance in the $\etaphi$ plane is larger than $\rs$. The effect
on the calculations is estimated by varying $\rs$ between $R$ and $2R$,
and amounts to $\sim 20\%$ for the measurements with $\r7$ in \cite{zeus2}.
For $\rr1$ the uncertainty is larger. Thus, this effect constitutes the major
source of uncertainty in the calculations. Unfortunately, the $\rs$
solution is limited only to NLO. Besides, the iterative cone algorithm
is not infrared safe beyond NLO. The best solution is to improve the
iterative cone algorithm or use the inclusive $\kt$ cluster algorithm \cite{kt}
instead.

\subsubsection{The $\kt$ cluster algorithm}
In the inclusive $\kt$ cluster algorithm \cite{kt}, jets are identified by
successively combining nearby pairs of particles until a jet is complete. The
$\kt$ algorithm allows a transparent translation of the theoretical jet
algorithm to the experimental set-up by avoiding the ambiguities related to
the merging and overlapping of jets and it is infrared safe to all orders.

\section{The structure of the photon}
The photon parton densities contain a hadronic part, which behaves like a
normal hadron, and a part called anomalous, which is the direct coupling of
the photon to a $q\bar q$ pair,

$$f_{i/\gamma}(x_{\gamma},\mu_F^2)=
f_{i/\gamma}^{\rm had}(x_{\gamma},\mu_F^2)+
f_{i/\gamma}^{\rm anom}(x_{\gamma},\mu_F^2).$$

At LO QCD, the photon structure function $\f2$ is proportional to the
densities for $q$ and $\bar q$,

$$\f2(x_{\gamma},\mu_F^2)=
\sum_qx_{\gamma}e_q^2[f_{q/\gamma}(x_{\gamma},\mu_F^2)+
f_{\bar q/\gamma}(x_{\gamma},\mu_F^2)].$$

The structure of the photon can be then studied by direct measurements of
$\f2$ in DIS $e\gamma$. A compilation of all the measurements of $\f2$ may be
found in \cite{nisius}. From these measurements, parametrisations of the photon
parton densities are determined \cite{do,dg,lac,gs,grv,acfgp,whit,sas} and used
in other processes to test the universality of the photon structure. However,
since no momentum sum rule applies to the photon parton densities it is
difficult to determine the gluon density from $\f2$ as in the case of a
hadron.

Measurements of cross sections for processes involving a large momentum
transfer (e.g. jets, prompt photons, high$-p_T$ hadrons, heavy quarks) are
compared to NLO calculations to test QCD and the available parametrisations of
the photon parton densities. These measurements may also be used in global
analyses to constrain the photon parton densities. Effective parton densities
can also be extracted from this type of measurements. The photon structure may
be studied as a function of the virtuality, from real to quasi-real to virtual
photons.

\section{Jet cross sections and QCD tests}
The measured dijet cross section $\spt$ in $\ga$ interactions from the
TOPAZ \cite{topaz}, AMY \cite{amy} and OPAL \cite{opal1,opal2} collaborations
have been compared to NLO QCD calculations \cite{klasen1,kleinwort}. The
calculations give a reasonable description of the data. These comparisons will
allow the possibility of discriminating between different parametrisations
of the photon parton densities when uncertainties in data and theory decrease.

The H1 collaboration has measured the double-differential inclusive dijet
cross section $d^2\sigma/d\xo d\log E_T^2$ \cite{h12}. These cross sections
are compared to NLO QCD calculations \cite{klasen1} to discriminate between
different parametrisations of the photon parton densities, and have been also
used to extract a LO effective parton density in the photon. The resulting
photon parton density exhibits an increase with the scale $p_T$ which is
compatible with the logarithmic increase predicted by QCD (due to the anomalous
term). The gluon density in the photon at LO has been extracted from
measurements of dijet cross sections \cite{h13} and high$-p_T$ inclusive
charged particle production \cite{h14}. The measurements are found to be
consistent to each other and with current LO parametrisations of the gluon
density in the photon. However, care must be taken when interpreting these
measurements since they were performed assuming perfect knowledge of the quark
densities in the photon. The gluon density extracted from the dijet cross
sections contains additional uncertainties coming from the underlying event,
absolute energy scale of the jets and hadronisation effects, and the density
extracted from the inclusive charged particle measurements suffers from
uncertainties of the fragmentation functions.

The underlying parton dynamics can be probed by measurements of the angular
distributions in two- and three-jet events.
The distribution of the angle between the jet-jet axis and the beam
direction in the dijet centre-of-mass system ($\cos\theta^*$) is
sensitive to the spin of the exchanged particle in two-body processes. The
dominant contribution in resolved processes proceeds via gluon exchange and the
cross section is proportional to $(1-\cost)^{-2}$. Direct processes
proceed via quark exchange and the cross section is proportional
to $(1-\cost)^{-1}$. The measured $\scost$ for resolved and direct
processes separated according to $\xo\lessgtr 0.75$ for dijet invariant masses
$\mj>23$~GeV \cite{zeus3} shows clear differences as $\cost\rightarrow 1$.
NLO QCD calculations \cite{baer} give a good description of the data.

At higher dijet invariant masses, the dijet cross sections as a function of
$\mj$ and $\cost$ can be used to identify new particles or resonances that
decay into two jets by deviations of the measurements with respect to the QCD
predictions \cite{zeus4}.

The angular distribution of the least energetic jet in three-jet events
probes the dynamics beyond LO. The distribution of the angle $\psi_3$,
the angle between the plane containing the highest energy jet and the
beam, and the plane containing the three jets, is different from pure
phase space and consistent with QCD predictions \cite{zeus5}.

\section{Prompt photon production}
Prompt photon production in $\gp$ interactions proceeds via direct
($q\gamma\rightarrow q\gamma$) and resolved ($qg\rightarrow q\gamma$)
processes. Therefore prompt photon production provides a means to study the
quark and gluon content of the proton and the photon. The advantage of prompt
photon measurements with respect to jet cross sections is that the photon
emerges directly from the hard interaction and is free from hadronisation
effects.

At LO QCD, the calculation of the prompt photon plus jet production cross
section is similar to that of jet production, except that one of the
final-state jets is replaced by a photon. NLO corrections to the prompt photon
production cross section are sizeable and the effect is most significant at
large pseudorapidity of the photon. At present there are two sets of NLO
calculations \cite{gordon,maria}. The calculations are performed using the
phase space slicing method which allows the implementation of photon isolation
cuts and jet algorithms as in the experiment.

The measurements of the cross section for prompt photon plus jet
production with photon isolation cuts \cite{zeus6} are sensitive to the quark
densities in the photon since they are done in a region of phase space where
the gluon content of the photon is suppressed and the proton parton densities
are well constrained. The NLO calculations \cite{gordon,maria} describe
reasonably well the data \cite{zeus6}. The calculations based on the
GRV-G HO \cite{grv} parametrisations of the photon parton densities provide a
better description of the data than the ones based on GS-G-96 HO \cite{gs}.

\section{Inclusive hadron production}
Inclusive hadron production at high$-p_T$ provides a test of QCD and is
sensitive to the photon parton densities. The detection of single mesons
and baryons allows also the study of fragmentation. The advantages with
respect to measurements of jet cross sections are that the ambiguities due to
jet algorithms are not present, and the uncertainties coming from the energy
scale and the underlying event are largely reduced. On the other hand,
additional information about the fragmentation functions is needed. NLO QCD
calculations for inclusive hadron production have been made in the $\ga$
\cite{kramer1} and the $\gp$ \cite{kramer2} regimes.

The ratio of the measured cross section for inclusive hadron production
in $\gp$ as a function of $p_T$ \cite{h14} to the NLO QCD calculations
\cite{kramer2} shows that the calculations give a good description of the
data. This is also observed for the measured cross section as a function of
$\eta$. However, even at NLO the uncertainty coming from the
renormalisation and factorisation scales in the calculations is quite
large.

In $\ga$ interactions, the NLO calculations \cite{kramer1} fail to describe
the large $p_T$ tail of the cross section measured by the MARK II \cite{mark}
and TASSO \cite{tasso} collaborations at a centre-of-mass energy of 29 and
33.1~GeV, respectively. At LEP energies, the NLO QCD calculations lie
significantly below the data for $W_{\ga}<30$~GeV for $d\sigma/dp_T$ and
$d\sigma/d|\eta|$ \cite{opal3}. The agreement with the data improves for
higher $W_{\ga}$.

\section{Heavy quark production}
Charm production in $\ga$ interactions proceeds via
$\ga\rightarrow c\bar c$ for direct processes and
$\gamma g\rightarrow c\bar c$ for single-resolved processes, which are
predicted to give comparable contributions at LEP energies; the
contribution from double-resolved processes is expected to be small.
Therefore, charm production in $\ga$ interactions is sensitive to the gluon
density in the photon. The comparison of the NLO QCD calculations for the total
cross section for charm production \cite{drees} to the measurement by the L3
collaboration \cite{l3} shows that direct processes alone cannot explain the
data and that a significant gluon content in the photon is required.

Charm production in $\gp$ interactions proceeds via
$\gamma g\rightarrow c\bar c$ for direct processes and
$gg\rightarrow c\bar c$ for resolved processes. Therefore, charm
production in $\gp$ interactions is sensitive to the gluon content in the
proton and in the photon. To study the gluon content in the proton, a
region of phase space is selected so as to reduce the resolved
contribution. Charm quarks have been tagged by reconstructing $D^*$ mesons.
The H1 collaboration has extracted the gluon density in the proton by
measuring $D^*$ production in the $\gp$ and deep inelastic $ep$ scattering
regimes \cite{h15}. Both measurements agree, supporting the universality of
the gluon density in the proton.

To study the photon parton densities, cross sections for $D^*$ production
are measured \cite{zeus7} in a region of phase space where the resolved
contribution is sizeable and then compared to NLO QCD calculations
\cite{cnlo1,cnlo2} based on different parametrisations of the photon parton
densities. The NLO QCD calculations of $d\sigma/dp_{T,D^*}$ in the massive
scheme \cite{cnlo1} lie below the data, and the calculations of
$d\sigma/d\eta_{D^*}$ fail to describe the measurements for $\eta_{D^*}>0$.
The NLO QCD calculations in the massless scheme \cite{cnlo2} based on the
GS-G-96 HO \cite{gs} parametrisations of the photon parton densities are
closest to the data; however, in these parametrisations the charm and u-quarks
contribute equally.

The comparison of the measured dijet cross section as a function of $\xo$ for
charm tagged events to NLO QCD calculations \cite{zeus7} shows an excess of
the data with respect to the calculations at low $\xo$. This excess may
constitute an evidence for the presence of charm in the photon, which is not
included in the calculations.

\section{Virtual photon structure}
It has been confirmed from measurements of $\f2$ in $e\gamma$
interactions and observation of resolved processes in $\gp$ interactions,
and of single- and double-resolved processes in $\ga$ interactions that
real photons have partonic structure. Virtual photons should also have
partonic structure. QCD predicts that the parton densities of virtual
photons become logarithmically suppressed as $P^2$ (the virtuality of
the probed photon) increases at fixed $\mu^2$ (the scale of the
interaction). If $\mu^2>P^2$, it is possible to ``resolve'' the photon.
As in the case of the real photon, the virtual-photon parton densities
contain a hadronic and an anomalous part. The first measurements of
the virtual photon structure function were made by PLUTO \cite{pluto}.

There exist two sets of parametrisations of the virtual-photon parton
densities \cite{sas2,grs1}. Both parametrisations extrapolate the real
photon parton densities to the virtual photon regime. They differ in the
treatment of the hadronic part: in \cite{sas2}, a fit is performed to a
coherent sum of the lowest-lying vector meson states $\rho$, $\omega$ and
$\phi$, whereas in \cite{grs1} the hadronic part is estimated by using the
pion parton densities.

The virtual photon structure can be studied in $\gamma^*\gamma^*$,
$\gamma^*\gamma$ and $\gamma^*p$ interactions by measurements of jet cross
sections. The cross section for jet production using virtual photons at LO QCD
is similar to that for real photons except that the photon parton densities are
replaced by those of the virtual photon. NLO QCD calculations for jet
production using virtual photons in $\gamma^*p$ \cite{potterep} and
$\gamma^*\gamma$ \cite{potteree} interactions have been recently completed.

The data from ZEUS on the ratio of the dijet cross sections for resolved and
direct processes separated according to $\xo\lessgtr 0.75$ as a function of
$P^2$ \cite{zeus8} have been compared to LO \cite{potterep,florian} and NLO
\cite{potterep} calculations for different parametrisations of the
virtual-photon parton densities. These measurements will allow the possibility
of discriminating between different parametrisations when the uncertainties in
data and theory decrease.

Recently, new parametrisations of the parton densities for all virtualities of
the photon have been presented \cite{grs2}. The H1 collaboration has measured
the effective parton density for the virtual photon at LO as a function of
$P^2$ \cite{h16}. The comparison of these measurements to the LO predictions
\cite{grs2} shows that the calculations give a reasonable description of the
data for $P^2<\mu^2\equiv p_T^2$ but fail when $P^2$ approaches $p_T^2$.

\section{Jet substructure}
Parton radiation and fragmentation may be studied via the jet shape,
which is the average fraction of the jet's transverse energy inside an
inner cone of radius $r$ concentric with the jet defining cone $R$,

$$
\psi(r)=\frac{1}{N_{jets}}\displaystyle\sum_{jets}
\frac{E_T(r)}{E_T(r=R)}.
$$

QCD predicts that at sufficiently high $\etjet$ the jet shape is driven
by gluon emission off the primary parton and that gluon jets are broader
than quark jets. $O(\alpha_s^3)$ QCD calculations for the reaction
$AB\rightarrow {\rm jet}+X$ give the lowest non-trivial order contribution to
the jet structure and, therefore, present a strong dependence on $\mu_F$,
$\mu_R$ and $\rs$.

The ZEUS collaboration has measured the jet shape for jets in $\gp$
interactions using an iterative cone algorithm \cite{zeus9}. The comparison of
these measurements to the calculations from \cite{klasen2} shows that even
though the calculations are made only up to the lowest non-trivial order, they
are able to describe the data with only one value of $\rs$ for
$\etjet>17$~GeV. When higher order corrections to the jet shape become
available, measurements with an improved iterative cone algorithm or the $\kt$
cluster algorithm \cite{juan} will allow very stringent tests of QCD. The
$\kt$ algorithm allows also the study of subjets, which are jet-like objects
within a jet \cite{kt} and constitute an independent test of the jet
substructure.

\section{Summary and outlook}
Significant progress in comparing measurements and NLO QCD calculations of
jet, prompt photon, high$-p_T$ hadron and heavy quark production has been
achieved in photon-induced processes. Experimental and theoretical
uncertainties have been reduced to help constrain the parton densities in the
real and virtual photon. Measurements of jet shapes and subjets will allow a
very stringent test of QCD when higher order corrections become available.

Finally, given all the current wealth
of measurements, an effort should be made towards a global analysis of
hard photon-induced processes to obtain a new generation of photon parton
densities.

\vspace{0.5cm}
{\bf Acknowledgements.} I would like to thank the organisers for
providing a warm atmosphere conducive to many physics discussions and a
well organised conference. Special thanks to \mbox{Prof. J. Terr\'on} and
L. Labarga for a critical reading of the manuscript.

\end{document}